\documentclass{iau}

\usepackage{amsmath}
\usepackage{graphicx}
\usepackage{multirow}
\newcommand{\nafe}{$\rm [Na/Fe]$}
\newcommand{\zh}{$\rm [Z/Fe]$}

\newcommand{\mgf}{$\rm Mg4780$}
\newcommand{\atio}{$\rm aTiO$}
\newcommand{\tioi}{$\rm TiO1$}

\newcommand{\tioii}{$\rm TiO2$}

\newcommand{\nad}{$\rm NaD$}
\newcommand{\naii}{$\rm NaI8190$}

\newcommand{\cai}{$\rm Ca1$}
\newcommand{\caii}{$\rm Ca2$}
\newcommand{\cnii}{$\rm CN2$}
\newcommand{\caiii}{$\rm Ca3$}
\newcommand{\cfs}{$\rm C4668$}
\newcommand{\mgi}{$\rm Mg1$}
\newcommand{\mgii}{$\rm Mg2$}

\newcommand{\feff}{$\rm Fe4531$}
\newcommand{\hbo}{$\rm H\beta_o$}

\newcommand{\mgb}{$\rm Mgb5177$}
\newcommand{\hgf}{$\rm H\gamma_F$}
\newcommand{\caf}{$\rm Ca4227$}

\begin{document}

\lefttitle{F. La Barbera, A. Vazdekis, A. Pasquali}
\righttitle{Chemical abundance ratios for the bulge of M31}

\jnlPage{1}{7}
\jnlDoiYr{2025}
\doival{10.1017/xxxxx}
\volno{395}
\pubYr{2025}
\journaltitle{Stellar populations in the Milky Way and beyond}

\aopheadtitle{Proceedings of the IAU Symposium}
\editors{J. Mel\'endez,  C. Chiappini, R. Schiavon \& M. Trevisan, eds.}

\title{Chemical abundance ratios for the bulge of M31}

\author{F. La Barbera$^{1}$, A. Vazdekis$^{2}$, A. Pasquali$^{3}$}
\affiliation{$^{1}$INAF-Osservatorio Astronomico di Capodimonte, sal. Moiariello
16, Napoli, 80131, Italy\\\email{francesco.labarbera@inaf.it}}
\affiliation{$^{2}$Instituto de Astrof\'\i sica de Canarias, Calle V\'\i a L\'actea s/n, E-38205
  La Laguna, Tenerife, Spain}
\affiliation{$^{3}$Astronomisches Rechen-Institut, Zentrum f\"ur Astronomie, Universit\"at Heidelberg, M\"onchhofstr. 12-14, D-69120 Heidelberg, Germany}

\begin{abstract}
We present abundance ratio estimates of individual elements, namely C, N, Na, and the so-called alpha elements, Mg, O, Si, Ca, and Ti, for the bulge of M31. The analysis is based on long-slit, high-quality, spectroscopy of the bulge, taken with the OSIRIS spectrograph at the Gran Telescopio CANARIAS (GTC). Abundance ratios, $\rm [X/Fe]$s, are inferred by comparing radially binned spectra of M31 with different state-of-the-art stellar population models, averaging out results from various methods, namely full-spectral, full-index, and line-strength fitting, respectively. For the bulk of the bulge, we find that O, N, and Na are significantly enhanced compared to Fe, with abundances of $\sim$0.3~dex, followed by C, Mg, and Si, with $\rm [X/Fe] \sim 0.2$~dex, and lastly Ti and Ca, mostly tracking Fe ($\rm [X/Fe] < 0.1$~dex), within the error bars. Performing the same analysis  on SDSS stacked spectra of early-type galaxies with different velocity dispersion, we find that the abundance pattern of the M31 bulge is very similar to that of most massive galaxies, supporting a scenario where most of the bulge formed in a fast and intense episode of star-formation.
\end{abstract}


\maketitle

\section{Introduction}

Studying the stellar population properties of stellar systems, namely their age, metallicity, the stellar IMF, and chemical abundance ratios is of paramount importance for our  understanding of the overall picture of galaxy formation and evolution. Abundance ratios are intimately connected to the high-mass end of the stellar IMF, as different elements are produced by stars with different masses, and because of different stellar lifetimes, they also inform us on the timescales of star-formation.
For most galaxies, having unresolved  stellar populations, the estimate of abundance ratios requires a detailed comparison of high S/N-ratio spectra with detailed predictions of stellar population models, accounting for the effect of varying abundance ratios. However, in order to provide reliable results, this technique needs to be extensively tested on stellar systems for which chemical compositions can also be constrained independently, based on resolved stellar population studies. Especially nowdays, thanks to the advent of JWST, and in the near future, with the E-ELT, such an opportunity is offered by nearby galaxies, such as the Andromeda galaxy (M31). 

Due to its proximity, the bulge of M31 gives us a unique opportunity to obtain high S/N spectroscopy at low observational cost, and apply the same techniques used to constrain the stellar population content of massive early-type galaxies (ETGs). Indeed, the stellar populations in the bulge are old, enhanced in alpha elements, and have super-solar metallicity in the center, similar to massive galaxies~(e.g.~\citealt[hereafter S10]{Saglia:2010}). The stellar population properties of the bulge  have been the subject of a long-standing debate, since the '70s. \citet{SpinTa:71} found evidence for a dwarf-dominated (bottom-heavy) IMF in the bulge, given the strong Na stellar absorption at $\sim 8200$~\AA , while~\citet{Faber:1972}, based on the first CO overtone at 2.3~$\mu$m, favoured a giants-dominated population. Nowadays, the bulge of M31 has been found to exhibit only a midly bottom-heavy IMF (not as heavy as the one suggested in the '70s), mostly confined to the innermost ($<10''$) regions~\citep[hereafter LB21]{LB:21}, which is also consistent with the bulge relatively low velocity dispersion ($\sim 150$~km/s; e.g.~S10). However, and surprisingly, detailed abundance ratios for the bulge of M31 remain poorly constrained. For instance, \citet{CvD12b} concluded that Na abundance is as high as 1~dex in the center of the bulge, in contrast to what observed in the bulge of the Milky-Way~\citep{Bensby:17} (but see~\citealt{Lec:2007}). Moreover, \citet{Z:15} claimed that different stellar population models produce very different predictions on the Na content of the M31 bulge, again favouring an extremely high \nafe\ value. 

In the present proceeding paper, we present a determination of individual abundance ratios for the bulge of M31, based on homogeneous, longslit, spectroscopy obtained with the OSIRIS spectrograph at the Gran Telescopio CANARIAS (GTC). Results are compared to those for SDSS stacked spectra of massive ETGs. We adopt a distance of 785~kpc from the MW to M31~\citep{McConnachie:2005}, implying a conversion scale of $\rm \sim 3.8$~pc/arcsec.

\section{Data and models}
\label{sec:datmod}
The bulge of M31 was observed on August and September 2017, with the OSIRIS instrument at the Nasmyth-B focus of GTC, at Roque de los Muchachos Observatory. Observations were carried out using a 7.4'-long slit of width 0.4'', with the U-, V-, R-, and I- R2500 grisms, resulting into a uniform spectral resolution of $\sim 16$~km/s. Further details on the observations and data reduction procedure can be found in LB21. For the present work, we use the same set of radially binned spectra as in LB21. Shortly, the 2D spectra of M31 are corrected for rotation velocity, and then a set of adaptively binned spectra is extracted, ensuring a  minimum S/N of 70 (per \AA ) in each bin. The binned spectra cover a spatial region out to a distance of $\sim 150$'' ($\sim 570$~pc) from the center of the bulge (see figure~1 of LB21).

We also use SDSS stacked spectra for low- and high- $\sigma$ ETGs in the local Universe ($\rm 0.05 \le z \le 0.095$), from~\citet{LB:13}. In particular, we use their stacked spectra with $\sigma=140$--$150$~km/s and $280$--$320$~km/s, respectively. We note that the low-$\sigma$ stack has the same velocity dispersion as the bulge of M31 ($\sigma \sim 150$~km/s, see S10), while the high-$\sigma$ includes the most massive ETGs.  The SDSS spectra are analyzed in the same way as M31 (see below).

We rely on two sets of stellar population models, i.e. EMILES~\citep{Vazdekis:12, Vazdekis:2016} and CvD18~\citep{CvD12a, CvD18} models. EMILES  simple stellar populations (SSPs) cover the spectral range from $0.35$ to $5 \, \mu$m, based on different empirical stellar libraries. The models are computed for two sets of (scaled-solar) theoretical isochrones, namely the ones of \citet{Padova00} (Padova00) and those of \citet{Pietrinferni04} (BaSTI); for different ages, from a few hundred Myr up to $\sim 14$\,Gyr; for different metallicities, \zh, up to supersolar values of $\sim 0.2$~dex (the exact value depending on the isochrones); and for different IMFs (see, e.g., LB21 for details). For the present work, we adopt a dedicated version of the models (hereafter Na--EMILES), where the SSPs are computed for a range of \nafe\ abundance ratios (see~\citealt{LB:17}). CvD18 models are an updated version of those of~\citet{CvD12a}. The models, covering the wavelength range 0.35--2.5~$\mu$m with the aid of empirical spectral libraries, are based on the MIST isochrones~\citep{Choi:2016, Dotter:2016}, and cover a range of ages (1--13.5~Gyr) and metallicities (from -1.5 to 0.2~dex). The SSPs are computed for different IMFs, using a three-segment power-law parametrization, with breaking points at 0.5 and 1~$\rm M_\odot$ (see CvD18 for details), respectively. The models also include a set of theoretical SSPs, computed for a Kroupa IMF, with varying abundance ratios for different elements, such as C, N, Na, and the alpha elements (Mg, O, Si, Ca, Ti). In the present work, we do not consider other elements, which are also varied in CvD18 models (e.g. Ba, Eu, etc...), as they have a tiny, sub one per cent effect, on galaxy spectra, and are much more difficult to constrain.

\section{Inference of abundance ratios}
We estimate abundance ratios using three different techniques:

\begin{description}
\item[]{\it Full spectral fitting (FSF)} - we adopt a similar approach as in~\citet{CGvD:2014}, fitting the spectra in wavelength space, over four wavelength ranges, i.e. 4000--4800~$\AA$,  4800--5800~$\AA$,  5800--6400~$\AA$, and 8000--8800~$\AA$. For each wavelength range, the fit is performed by normalizing both spectra  and models with multiplicative polynomials of degree $\delta \lambda /100$, where $\delta \lambda$ is the size of the wavelength interval. The fitting is performed with a dedicated Fortran77 code, using a Levenberg–Marquardt $\chi^2$-minimization algorithm. Initial conditions on fitting paramaters (namely, ages, metallicities, IMF slope, and abundance ratios) are varied over a discrete grid of input values, to ensure that a global minimum is always found. Emission lines in the spectra are removed by fitting Gaussian functions simultaneously with the stellar component. 
\item[]{\it Full Index Fitting (FIF)} - we adopt the same approach as in~\citet{NMN:19}, by performing spectral fitting only on the regions corresponding to a well-selected set of spectral indices. We select a large sample of optical and NIR indices, as in table~1 of LB21 (i.e. \hbo,  \hgf, \tioi, \tioii, \atio, \mgf, \nad, \naii, \cai, \caii, \caiii, \caf, \mgi, \mgii, \cfs, \cnii, \mgb, \feff; see their case K), adding  other Lick-like indices, such as Fe4383, Fe5015, Fe5270, Fe5335, $\rm H\delta_F$, and G4300. For each index, both observed  and model spectra are normalized by the line passing through the corresponding pseudocontinua, and the $\chi^2$ minimization is performed only on the index passbands.
    \item[]{\it Line-strength fitting (LSF)} - we adopt the same approach as in~LB21 (and references therein), by fitting index line-strengths, rather than spectra in wavelength space. We consider the same set of indices as for the FIF approach.
\end{description}
For each method, the fits are repeated for both EMILES or CvD18 models, assuming either a single SSP (1SSP) or an older+younger (2SSP) stellar component. For the FIF and LSF approaches, the input spectra are preliminary cleaned up for emission lines, by subtracting the best-fitting Gaussian functions from the FSF. ISM contamination in the NaD line is removed as detailed in LB21. Best-fitting coefficients include age and metallicity for the single or the two SSPs, the mass fraction of the second component, if present, the IMF slopes (a single slope for EMILES models, where a low-mass tapered IMF parametrization is adopted; or two slopes, up to 0.5~$\rm M_\odot$, and from 0.5 to 1~$\rm M_\odot$,  in the case of CvD18 models), and abundance ratios for C, N, Na, and the alpha elements Mg, O, Si, Ca, and Ti.
In all cases, CvD18 models are used to compute spectral responses to varying individual abundance ratios, but for the case of EMILES and \nafe\ abundances, where the responses are computed with Na--EMILES models (see Sec.~\ref{sec:datmod}). We note that in the case of ``perfect'' models, i.e. without systematic deviations from observed spectra, all three approaches (FSF, FIF, LSF) would provide fully consistent results within the statistical uncertainties. If this were the case, FSF would be preferable, as it provides lower statistical uncertainties on best-fitting parameters (e.g.~\citealt{vanDokkum:2017}). However, in practice, because of systematic deviations between data and models, and because of uncertainties on the responses to varying abundances (based on theoretical atmosphere models), different fitting methods do not necessarely provide consistent results, making it not so obvious the best approach to adopt. For this reason, in the present work, we just average out results from different methods and models, and use the corresponding standard deviations as an estimate of the uncertainties on best-fitting parameters (i.e. abundance ratios).

\begin{figure}
\centering
  \includegraphics[scale=.6]{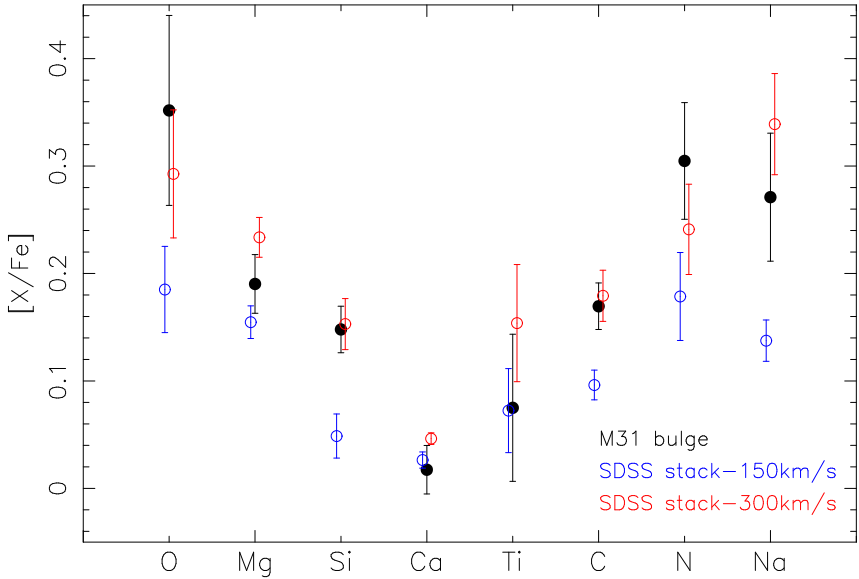}
  \caption{Abundance ratios, $\rm [X/Fe]$s, for different chemical elements (see labels on the x-axis) are plotted for the bulge of M31 (black dots), and the SDSS stacked spectra  of ETGs with low-  and high- velocity dispersion ($\sim$150 and $\sim$300~km/s, respectively; see labels on the lower--right, and Sec.~\ref{sec:datmod}). Error bars correspond to 1~sigma uncertainties, accouting for variations among results based on different fitting methods and stellar population models. Note that the figure plots, from left to right, alpha elements first, and then the other elements (C, N, and Na), both groups in order of increasing atomic number. }
  \label{fig:xfe}
\end{figure}

  \section{Results and Conclusions}
  Fig.~\ref{fig:xfe} shows average values of  best-fitting abundance ratios for the bulge of M31, compared to those of SDSS stacked spectra for low- and high-$\sigma$ ETGs ($\sim $150 and 300~km/s, respectively). For the bulge of M31, we average results for all radial bins in the region outside $10$'', where most spectral indices, as well as inferred  stellar population properties (including the stellar IMF), exhibit flat radial trends (see LB21). This region corresponds to the bulk of the bulge luminosity.  The Figure shows that O, N, and Na are all significantly enhanced in the bulge of M31, with abundance ratios of $\sim 0.3$~dex, followed by C, Mg, and Si, having $\rm [X/Fe] \sim $0.2~dex, and finally Ca and Ti, tracking Fe (with $\rm [X/Fe] < $0.1~dex) within the error bars. The largest error bars are for O and Ti, whose abundance estimates show large variations among different techniques and models. We note that the abundances of C and O are consistent with those derived from the analysis of CO absorption lines in the H- and K-band spectral range~\citep{LB:24}, while the abundances of Mg and Na are consistent with those derived from LB21, based on the same data. The Mg abundance agrees with previous estimates of $\rm [\alpha /Fe]$ from S10.
  
Interestingly, Fig.~\ref{fig:xfe} shows that the  bulge of M31 does not show the same abundance pattern as ETGs with similar velocity dispersion ($\sim 150$~km/s; see black and blue symbols in the Figure). All the elements tend to be more enhanced in the bulge, but for those (Ca and Ti) that mostly track Fe. On the contrary,   the abundance pattern of the bulge is very similar to that of most massive ETGs, as seen by comparing black and red symbols in Fig.~\ref{fig:xfe}.  This points to a scenario where most of the M31 bulge formed in a quick and intense episode of star-formation at high redshift, in agreement with what predicted by chemical evolution models for this system~\citep{MU:2015}.

\begin{acknowledgements}
F.L.B. and A.P. acknowledge support from INAF mini-grant 1.05.23.04.01.
\end{acknowledgements}

\end{document}